\documentclass[aps,twocolumn,amssymb,showpacs,amsmath,preprintnumbers]{revtex4}

\usepackage{graphicx}

\begin{document}

\title{ Formation of a sonic horizon in isotropically expanding Bose-Einstein condensates}

\author{Yasunari Kurita}
\affiliation{Osaka City University Advanced Mathematical Institute, Osaka
558-8585, Japan}

\author{Takao Morinari}
\affiliation{Yukawa Institute for Theoretical Physics, Kyoto
        University, Kyoto 606-8502, Japan}

\preprint{YITP-07-16}
\preprint{OCU-PHYS-264}
\preprint{AP-GR-40}

\begin{abstract}
We propose a simple experiment to create a sonic horizon
in isotropically trapped cold atoms within currently available experimental techniques.
Numerical simulation of the Gross-Pitaevskii equation shows
that the sonic horizon should appear by making the condensate expand.
The expansion is triggered by changing the interaction which can be 
controlled by the Feshbach resonance in real experiments.
The sonic horizon is shown to be quasi-static for sufficiently
strong interaction or large number of atoms. 
The characteristic temperature that is associated with 
particle emission from the horizon, which corresponds to 
the Hawking temperature in an ideal situation,
is estimated to be a few nK.
\end{abstract}

\pacs{03.75.Kk, 03.75.Hh, 04.62.+v, 05.30.Jp}

\maketitle

\section{Introduction}

For the exploration of cosmology and gravitational physics, 
it is necessary to have a deep understanding of quantum filed theory in curved spacetime: 
It is widely believed that everything except for the spacetime itself should 
originate from quantum fluctuations in the early Universe. 
Quantum effects on curved spacetime, such as the Hawking radiation, 
give us theoretical support for black hole thermodynamics. 
However, it is extremely hard to verify such quantum effects experimentally. 
For instance, the Hawking radiation is thermal radiation emitted 
from a dynamically formed stationary black hole \cite{hawking}. 
However, the characteristic temperature of the thermal radiation, 
the Hawking temperature, is on the order of several tens of nanokelvins at most, 
which is much lower than the cosmic microwave background radiation temperature. 
So detecting thermal radiation from a real black hole is almost impossible.

One way to circumvent this difficulty is to make use of 
artificial black holes \cite{novello2002}\cite{Barcelo:2005fc}.
Unruh showed in his seminal paper\cite{unruh81} that 
excitations in a supersonic flow corresponds to 
a scalar field equation on a curved spacetime including a horizon. 
Since the phenomenon of the Hawking radiation can be separated from 
gravitational physics, 
it is possible to detect the corresponding phenomenon 
in a fluid system with sonic horizon \cite{unruh81}. 
The basic idea is to identify fluid flow with curved spacetime and 
excitation modes with fields on the curved spacetime. 
A black hole event horizon corresponds to a sonic horizon in a fluid. 
For the purpose of investigating the quantum effects, 
a quantum fluid should be considered. 
As such a quantum fluid, 
Bose-Einstein condensates (BEC) 
in trapped cold atoms \cite{Anderson-95,Ketterle95} are 
one of the most suitable systems \cite{Garay-PRL}-\cite{Barcelo:2000tg}.
A crucial advantage is that one can control scattering length
between atoms by making use of the Feshbach resonance\cite{Feshbach}.
In fact, that experimental technique was used in observing 
jets and bursts in a collapsing condensate, 
which is called ``Bose-Novae" \cite{Bose-Novae}. 
An remarkable explanation of burst and jet phenomena in Bose-Novae was proposed 
in~\cite{Calzatta0207}\cite{Calzatta0208}, based on quantum field theory of particle creation and structure formation
in cosmological spacetime. 

In order to verify the Hawking effect in fluid analogy, 
it is necessary to create a stationary sonic horizon 
because it is a phenomenon on a dynamically formed stationary black hole. 
Although several possibilities have been discussed so far \cite{Garay-PRL,Garay:2000jj}, 
it seems difficult to realize exactly stationary sonic horizon in cold atoms. 
However, if one can make a quasi-static horizon for high frequency modes, 
particle emission from the horizon is also expected.
In this paper, 
we numerically demonstrate that a quasi-static horizon is realized
without introducing new experimental techniques 
beyond currently available ones.
We consider an expanding BEC driven by a sudden change of the interaction.
Numerically solving the Gross-Pitaevskii (GP) equation,
we show that a quasi-static horizon appear.

We note that there have been made great efforts 
to create cosmological geometry using 
expanding BEC\cite{Barcero02}-\cite{Weinfurtner04}. 
In these papers, the analogue models with specific cosmological metrics such as 
Friedmann-Robertson-Walker (FRW) metric or de Sitter metric were discussed and 
the effects of particle creation in these cosmological spacetimes were investigated. 
However, in this paper, 
we do not intend to obtain any cosmological analogue model with well-known analytic metric. 
But we try to obtain dynamically formed quasi-static sonic horizon. 
Furthermore, the sonic horizon should be formed in hydrodynamic regime of the condensate because
the spacetime analogy is only valid in such regime. 
The appearance of a horizon due to expansion of a condensate was 
noticed in the 
previous works, for example in \cite{Barcero03}, and its formation itself is not surprising. 
But it is non-trivial whether the condensate flow at the horizon is in the hydrodynamic regime, or not. 
In this paper, we show that the quasi-static sonic horizon will appear in the hydrodynamic regime of the condensate
by changing the atomic interaction instantaneously.

 \section{Analogue spacetime in BEC}

In the coherent state path integral formulation, the action of bosons
is given by
\begin{eqnarray}
S=
\int dt \int d^3 {\bf r}
\bigg(i\hbar \bar{\phi}\partial_{t} \phi
   -\frac{\hbar^2}{2m}\nabla\bar{\phi}\cdot\nabla \phi 
 - V_{\rm ext}\bar{\phi}\phi \nonumber \\
\hspace{3cm} -  \frac{1}{2}U_0 (\bar{\phi}\phi)^2 \bigg),
\end{eqnarray}
where 
$V_{\rm ext}$ is the confining potential 
and
$U_0={4\pi \hbar^2 a}/{m}$ with $a$ the $s$-wave scattering length.
For $\phi$ and $V_{\rm ext}$, spatial and time dependences 
are implicit.
The saddle point equation for this action leads to
the GP equation:
\begin{eqnarray}
  i\hbar \partial_{t} \Psi = 
        \left( -\frac{\hbar^2}{2m} \nabla^2+V_{\rm ext}
            + U_0 |\Psi|^2\right)\Psi.
 \label{eq:GP}
\end{eqnarray}
This GP equation governs the dynamics of the condensate whose 
order parameter is given by $\Psi$.

Now we consider hydrodynamical approximation.
We denote the bosonic field $\phi$ as 
$\phi = \sqrt{\rho_0+\rho}e^{i(\varphi_0+\varphi)}$,
where $\sqrt{\rho_0}$ and $\varphi_0$ are the amplitude
and the phase of $\Psi$, respectively.
(Namely, $\Psi = \sqrt{\rho_0} \exp(i\varphi_0)$.)
The fields $\rho$ and $\varphi$ describe the
non-condensate part of the bosonic field. 
If the density gradient is sufficiently smooth over the scale determined by 
the local healing length 
$\xi ({\bf r}, t) \equiv \hbar/(2m\rho_0 U_0)^{1/2}$, 
or, in other words if the conditions,
\begin{eqnarray}
|\xi \nabla \rho_0/\rho_0  |^2 \ll 1\quad  \mbox{and} \quad  |\xi \nabla \rho/\rho  |^2 \ll 1,
\label{eq:condition-healing}
\end{eqnarray}
are satisfied, 
hydrodynamical approximation is justified.
(The condition (\ref{eq:condition-healing}) shall be examined later.)
Under the above condition, the equation for $\rho$ is 
\begin{eqnarray}
\rho =-\hbar (\dot{\varphi} +{\bf v_0} \cdot \nabla \varphi)/U_0, 
\label{eq:rho-varphi}
\end{eqnarray}
where ${\bf v_0}=\frac{\hbar}{m}\nabla \varphi_0$ 
is the background fluid velocity, 
and the effective action for $\varphi$ is 
\begin{eqnarray} 
S_{\rm eff}=
\int dt \int d^3 {\bf r}
\frac{\hbar^2}{2U_0} \left[(\dot{\varphi}+{\bf{v}_0} \cdot \nabla \varphi )^2 
-\frac{\rho_0 U_0}{m}
\left(\nabla \varphi\right )^2 \right].
\end{eqnarray}
Taking variation with respect to $\varphi$, we find that
the field equation for $\varphi$ has the form of a propagating wave 
equation. 
Also, the equation for the field $\varphi$ can be expressed as
 $\partial_{\mu}(\sqrt{-g}g^{\mu\nu} \partial_{\nu})\varphi =0$, 
where $g^{\mu \nu}$ is the inverse 
of the following matrix:
\begin{eqnarray}
g_{\mu\nu} \propto 
\left(
 \begin{array}{cc}
  -(c_s^2-{\bf v_0}^2) & -{\bf v_0} \\
  -{\bf v_0} & {\bf 1} 
 \end{array} \right),
 \label{eq:effectivemetric} 
\end{eqnarray}
with $c_s =\sqrt{\rho_0 U_0/m}$ and
$g={\mbox{det}}g_{\mu\nu}$.
Thus, the equation is equivalent to
an equation for a massless field on a curved spacetime determined by the metric 
(\ref{eq:effectivemetric}) with $c_s$ the speed of "light." 
Note that in order to interpret the quantity $c_s$ as a velocity, 
$U_0$ must be positive because, for negative $U_0$, $c_s$ becomes pure imaginary. 
Hereafter we consider positive $U_0$, which leads to
an effective spacetime with Lorentzian signature.

For the excitation modes of $\varphi$ whose frequencies, say $\omega$,
are much higher than the frequency $\omega_{\rm BEC}$,
which is associated with the condensate motion, the condensate will be quasi-static. 
(For moderate changes of the interaction, $\omega_{\rm BEC}$ turns out to be 
the trapping harmonic potential frequency $\omega_{ho}$,
as shall be discussed below .)
The analogy between fields 
on the curved spacetime and excitation modes on the fluid flow is meaningful 
only when the conditions (\ref{eq:condition-healing}) are satisfied.
The latter condition in Eq.(\ref{eq:condition-healing}) turns out to be 
$\omega^2 \ll (c_s/\xi)^2$, by using Eq.(\ref{eq:rho-varphi}).
Thus, the frequency $\omega$ has an upper limit.
The former condition in Eq.(\ref{eq:condition-healing}) is satisfied 
in the regions far from the edge of the condensate.
(In contrast, if one is very close to the edge, 
zero-point oscillations become dominant,
and so the former condition in (\ref{eq:condition-healing})
is not satisfied.)
If there exists intermediate region for $\omega$ of 
\begin{eqnarray}
  \omega_{\rm BEC} \ll \omega  \ll c_s/\xi,
  \label{eq:intermediate}
\end{eqnarray}
then the hydrodynamical approximation is justified and 
the condensate is quasi-static for excitation modes.
Note that those modes are associated with 
particle emission from the horizon 
if the hydrodynamical flow has a dynamically formed sonic horizon.
The necessary condition for the existence of the intermediate region (\ref{eq:intermediate}) is
\begin{eqnarray}
\frac{c_s}{\xi \omega_{\rm BEC}} \gg 1. 
\label{eq:cond-fluid-static}
\end{eqnarray}
In the following, we mainly consider condensate satisfying the above 
condition.

\section{Formation of sonic horizon}
Now we investigate sonic horizon formation in an expanding BEC trapped in 
isotropic harmonic potential, 
$V_{\rm ext} =m \omega_{ho}^2 r^2/2$, where $r$ is the radial coordinate. 
Initially, we set the condensate in a ground state with an initial atomic interaction $a_i$.
At $t=0$, the atomic interaction is changed suddenly from $a_{i}$ to $a_{f}(>a_{i})$, 
which makes the condensate expand.
Then, formation of sonic horizon can be expected. 
The reason is as follows:
The sound velocity is proportional to square root of the condensate
density and a decreasing function of $r$. 
In contrast, the fluid velocity is an increasing function of $r$ 
and the condensate expands fast around its edge 
whereas $v_0 ({\bf r}={\bf 0}) = 0$ due to the boundary condition.
Therefore, at an intermediate radius, $v_0$ exceeds $c_s$ and the fluid flow is transonic.
It has a surface satisfying $c_s=|v_0|$ which is called a sonic horizon. 
We should note that the sonic horizon corresponds to a horizon in the analogue spacetime defined by
the metric (\ref{eq:effectivemetric}).
We also note that the subsonic region is around the center of the condensate and 
inside of the sonic horizon. 

In general, if a fluid has a static sonic horizon and 
a proper quantum state for an excitation field is realized, 
then it is theoretically predicted that the horizon will emit thermal radiation of the quantum field. 
As will be discussed in Appendix \ref{particle-creation}, 
if the sonic horizon in the expanding condensate is quasi-static for the field $\varphi$, 
the horizon will emit thermal radiation into the center of the condensate.
The temperature characterizing the thermal emission (Hawking temperature) 
is given by the following formula: 
\begin{eqnarray} 
T_{\rm pc} =\frac{\hbar }{2\pi k_{\rm B}}\partial_r (v_0-c_s)|_{r_{\rm H}},
\label{eq:Hawking-formula}
\end{eqnarray} 
where $r_{\rm H}$ is the horizon radius and $k_{\rm B}$ is the Boltzmann's constant\cite{unruh81}\cite{Jacobson03}.
From the above expression, it is found that 
the Hawking temperature is determined by gradient of fluid and sound velocity at the horizon.
Thus, it is important to investigate the velocity gradients at the horizon.
In the derivation of the formula (\ref{eq:Hawking-formula}), it is assumed that the dynamically 
formed horizon is static, but in actual experiments, this assumption is not satisfied exactly.
Therefore, the spectrum of the particle is not fully 
given by the single Planck's distribution function, but rather
given by a superposition of the Planck's distribution functions 
with slightly different temperatures.
Even if this is the case, the energy scale of the particle creation 
emitted from the dynamically formed horizon is on the order of $T_{\rm pc}$.

We have simulated the expansion of the condensate by solving numerically (using the Crank-Nicolson scheme)
the time-dependent GP equation.
The initial ground-state wave function is obtained by solving the GP equation 
using the steepest descent method for an initial s-wave scattering length $a_{i}$ 
and the number of atoms $N$. 
We have computed $c_s$ and the radial velocity of the condensate via
\begin{eqnarray}
c_s &=& \sqrt{(\Psi^*\Psi) U_0/m }, \\
v_0 &=& \hbar \left[ \Psi^* \partial_r \Psi
- \left( \partial_r \Psi^* \right) \Psi \right]/
(2mi|\Psi|^2),
\end{eqnarray}
and searched for parameter sets leading to $|v_0|>c_s$.

 \begin{figure}
   \begin{center}
    \includegraphics[width=8cm]{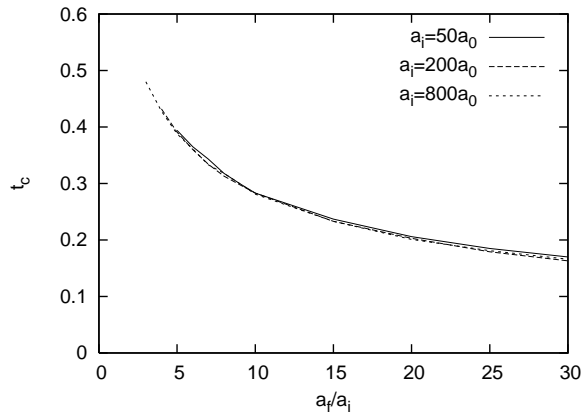}
   \end{center}
   \caption{ \label{fig:tc}
 Times $t_c$ in each simulation are shown in units of $\omega_{ho}^{-1}$.
 The horizontal axis is the ratio of $a_f$ to $a_i$.
    }
 \end{figure}

In the following, we assume that the condensate consists of $N=10^5$ Rb atoms. 
(The values of atomic interaction given below are those in the case of $N=10^5$.
If $N= 10^5/n$ with an integer $n$, then $a_i$ and $a_f$ should be multiplied by $n$.)
The initial atomic interaction is assumed to be $a_i=50 a_0,\ 200a_0$ and $800 a_0$ 
where $a_0=0.53 \times 10^{-10}$m is the Bohr radius. 
The following change of the atomic interaction has been simulated: 
$a_f/a_i = 2,3,4,5,6,7,8,9,10,15,20,25,30$.

Just after $t=0$, the condensate begins to expand in the trapping potential 
and the expansion is accelerated for a while.
At some time, say $t=t_c$, the expansion turns to be decelerated. 
Figure \ref{fig:tc} shows $t_c$ as a function of $a_f/a_i$. 
It is seen that $t_c$ does not depend on the initial strength of the interaction. 
For $\tilde{t}:=\omega_{ho} t < \pi/2$, the condensate continues to expand, and 
at $\tilde{t} \simeq \pi/2$, the condensate starts to collapse. 
Therefore, we turn off the trapping potential at $\tilde{t}=\pi/4$ and 
make the condensate expand freely in order to keep the horizon for a while. 

\begin{figure}
 \begin{center}
    \includegraphics[width=8cm]{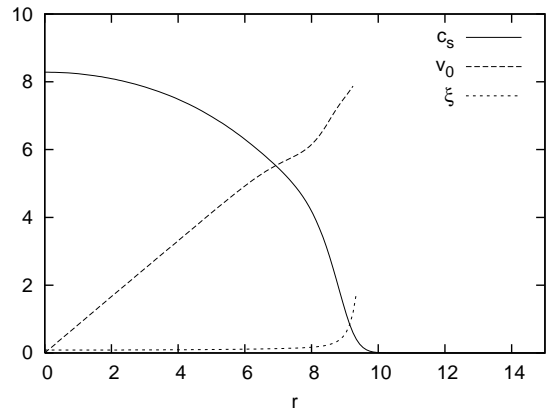}	
 \end{center}
 \caption{ \label{fig:cvxf1}
 Sound velocity $c_s$(solid line) and 
 the fluid velocity $v_0$(dashed line) 
 versus $r$ at $\tilde{t}=0.4$ in the case of $a_i=200a_0$ and $a_f =5a_i$ are shown
 in units of $(\hbar \omega_{ho}/m)^{1/2}$.
 The healing length $\xi$(dotted line) is shown 
 as well in units of 
 $a_{ho}=\left(\hbar/m\omega_{ho}\right)^{1/2}$.
 }
 \end{figure}

\begin{figure}
 \begin{center}
    \includegraphics[width=8cm]{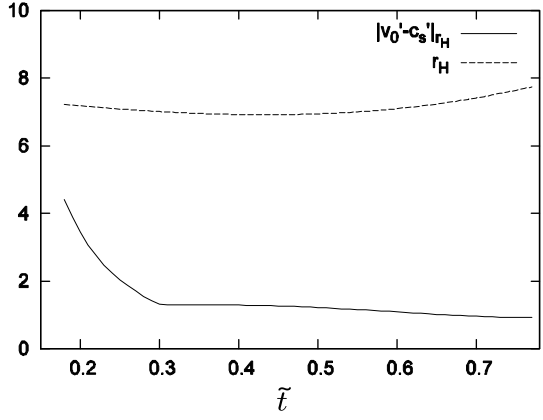}
 \end{center}
 \caption{ \label{fig:hp1-vcd1}
 Time dependence of 
 $\partial_r (v_0-c_s)_{r_H}$ (solid line) 
 in units of $\omega_{ho}$ and
 the position of the sonic horizon $r_H$ (dashed line)
 in units of 
 $a_{ho}=\left(\hbar/m\omega_{ho}\right)^{1/2}$ in the case of $a_i=200a_0$ and $a_f =5a_i$.
 }
 \end{figure}

As far as we have investigated, sonic horizon always appears in the sence of the surface 
where $v_0$ exceeds $c_s$. 
As an example of a sonic horizon, 
we show Fig. \ref{fig:cvxf1} which is the snapshot at $\tilde{t}=0.4$ 
in the case of $a_i=200a_0$ and $a_f =5a_i$.
We see that, around $r=7a_{ho}$, the fluid velocity exceeds the sound velocity, and the sonic horizon exists there.
In this case, we find that, at $\tilde{t} \equiv \omega_{ho} t=0.11$, the horizon appears.
Fig. \ref{fig:hp1-vcd1} shows the time dependence of
the radius of the horizon, say $r_H$, and the velocity gradient at the horizon 
$\partial_r (v_0-c_s)|_{r_H}$.

\begin{figure}
 \begin{center}
    \includegraphics[width=8cm]{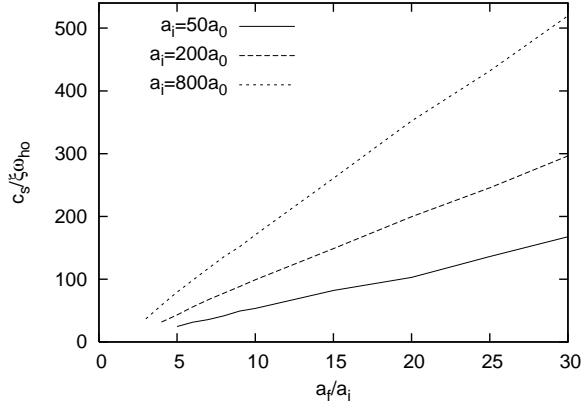}
 \end{center}
 \caption{ \label{fig:Xh}
  $c_s /\xi \omega_{ho}$ as a function of $a_f/a_i$ is shown for each initial scattering length.  
 }
 \end{figure}

\begin{figure}
 \begin{center}
    \includegraphics[width=8cm]{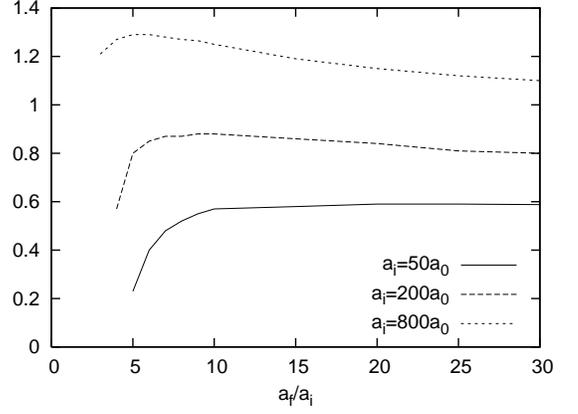}
 \end{center}
 \caption{ \label{fig:hlt}
 Horizon life time as a function of $a_f/a_i$ is shown in units of $\omega_{ho}^{-1}$. 
 }
 \end{figure}

If we keep the trapping potential for a long time, 
an oscillating behavior of the condensate is observed. 
The period of the oscillation is about $\pi$ in units of $\omega_{ho}^{-1}$ and 
we find $\omega_{\rm BEC} \simeq \omega_{ho}$,
within a moderate change of the interaction.
This oscillation is just like an oscillation of a droplet confined in a harmonic potential.
Therefore, the condition (\ref{eq:cond-fluid-static}) can be rewritten as
\begin{eqnarray}
 \frac{c_s}{\xi \omega_{ho} }\gg 1.
\label{eq:cond-fluid-static2}
\end{eqnarray}
Now, we are interested in sonic horizon where (\ref{eq:cond-fluid-static2}) is satisfied.
The intermediate region (\ref{eq:intermediate}) exists if,
for example, the following inequality is satisfied:
\begin{eqnarray} 
\frac{c_s}{\xi \omega_{ho} } \ge 22.5.
\label{cond-threshold} 
\end{eqnarray} 
For this choice of the lower bound, there exists a region 
of $\omega$ satisfying both conditions of 
$\omega \ge 10\omega_{ho} $ and $ (c_s /\omega \xi )^2 \ge 5 $.
The condition (\ref{cond-threshold}) ensures hydrodynamic flow and quasi-static nature of the condensate. 
Fig. \ref{fig:Xh} shows $c_s/\xi \omega_{ho}$ as a function of $a_f/a_i$.
We define horizon life time as the time interval during which the condition (\ref{cond-threshold}) 
continues to be satisfied at the horizon. 
The horizon life time is shown in Fig. \ref{fig:hlt}.  
As far as we have investigated, 
the condensate flow satisfying (\ref{cond-threshold}) at sonic horizon appears only 
when $a_f \geq 5a_i$ for $a_i = 50a_0$, $a_f \geq 4a_i$ for $a_i=200a_0$ and $a_f \geq 3a_i$ for $a_i=800a_0$.

\begin{figure}
 \begin{center}
    \includegraphics[width=8cm]{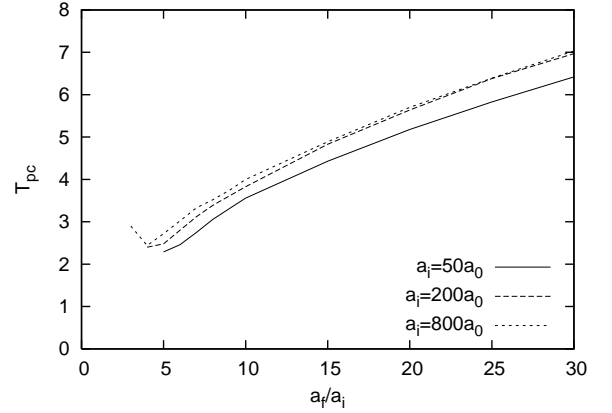}
 \end{center}
 \caption{ \label{fig:TH}
 Hawking temperature at $t=t_c$ in units of nK. In the evaluation, we assume $\omega_{ho}=1400$ Hz. 
 }
 \end{figure}

\begin{figure}
 \begin{center}
    \includegraphics[width=8cm]{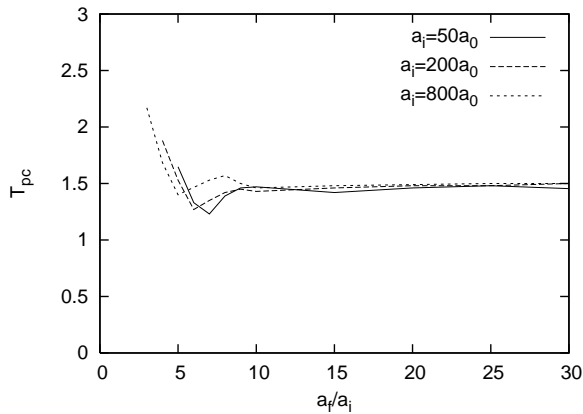}
 \end{center}
 \caption{ \label{fig:TH2}
 Hawking temperature at $\tilde{t}=0.79$ in units of nK, 
 when just after the trapping potential is turned off. 
 In the evaluation, we assume $\omega_{ho}=1400$ Hz.   }
 \end{figure}

The Hawking temperature at $t=t_c$ and $\tilde{t} =0.79$ 
(just after turning off the trapping potential) are shown in Fig. \ref{fig:TH} and
Fig. \ref{fig:TH2}, respectively. 
In the evaluation, we assume the frequency $\omega_{ho} =$ 1400 Hz. 
The Hawking temperature at $t=t_c$ depends on the ratio $a_f/a_i$ almost linearly.
In contrast, for $a_f/a_i\geq 9$, 
the Hawking temperature at $\tilde{t}=0.79$
does not depend on the ratio so much.
From the simulations, the temperature is expected to be a few nK.

For this spherically symmetric trap, one may concern the three-body recombination loss of condensed atoms.
Now, we check the effect of three-body losses for the given peak density.
This effects may be taken into account by incorporating the imaginary term describing the 
inelastic process in the GP equation ~\cite{Kagan1998}
\begin{eqnarray*}
i\hbar {\partial_t} \Psi = \left( -\frac{\hbar^2}{2m} \nabla^2 +V_{ext}
+U_0 |\Psi|^2 \right) \Psi -\frac{i\hbar }{2} K_3 |\Psi|^4 \Psi,
\end{eqnarray*}
where $K_3$ denotes three-body recombination loss-rate coefficient.
Then, the three-body loss is proportional to the cube of the atomic density
\begin{eqnarray*}
\frac{\partial}{\partial t} \int |\Psi|^2 d^3r = -K_3 \int |\Psi|^6 d^3 r,
\end{eqnarray*}
which implies that the three-body loss rate is given by $R_3 \equiv K_3 \int |\Psi|^6d^3 r/ \int |\Psi|^2 d^3r$.
For the value of $K_3$,
we assume $K_3 = 2\times 10^{-28} $cm$^6/$s, 
according to \cite{Saito-Ueda2002}.
Of course, high atomic density causes many inelastic processes and gives high atomic loss rate.
In our numerical simulation, the upper limit of the loss rate can be estimated by use of the peak density as
$R_3 \leq 3 \times 10  \mbox{s}^{-1}$, where
the total atomic number was set to be $N=\int |\Psi|^2 d^3{\bf r} = 10^5$. 
Then, the three-body loss can be ignored because we consider the time scale of $\leq 10$ ms.

\begin{figure}
 \begin{center}
    \includegraphics[width=8cm]{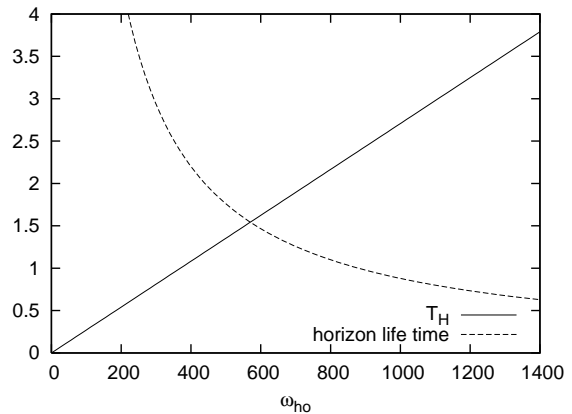}
 \end{center}
 \caption{ \label{fig:omega-dep}
 $\omega_{ho}$-dependence of $T_H$ and horizon lifetime in the 
 case of $a_i=200a_0$ and $a_f =10a_i$.
 The Hawking temperature is shown in units of nK and horizon lifetime is in units of $\omega_{ho}^{-1}$.
  }
 \end{figure}

In the above evaluation for Hawking temperature, horizon lifetime and $R_3$, 
we have assumed that the trapping frequency is $\omega_{ho} =$ 1400 Hz.
Note that $\omega_{ho}$ is the energy scale of the system.
Therefore, a large value of $\omega_{ho}$ is plausible to
increase the characteristic temperature for the particle emission,
though the time evolution process becomes rapid for large $\omega_{ho}$.
If lower frequency is assumed, 
lower temperature, longer horizon lifetime and fewer three-body loss rate would be expected. 
As an example, 
Fig. \ref{fig:omega-dep} shows $\omega_{ho}$-dependence of the Hawking temperature and the horizon lifetime
in the case of $a_i=200a_0$ and $a_f =10a_i$.

\section{Bogoliubov spectrum}
In the above numerical simulations, we assume there is no dynamical instability.
Now, we check whether there is dynamical instability or not, within 
Gaussian approximation. 
For that purpose, we study the Bogoliubov-de Gennes equations:
the second quantized field equations 
for the excitation fields $\delta \phi$ and $\overline{\delta \phi}$
are given by
\begin{eqnarray*}
i\hbar \partial_t \delta \phi &=& \left(
-\frac{\hbar^2}{2m} \nabla^2 +V_{\rm ext}
+2U_0 |\Psi|^2 \right)\delta\phi  + U_0 \Psi^2 \overline{\delta \phi}, 
\\
-i\hbar \partial_t \overline{\delta \phi} &=& \left(
-\frac{\hbar^2}{2m} \nabla^2 +V_{\rm ext} 
+2U_0 |\Psi|^2 \right)
\overline{\delta \phi}
+ U_0 \left(\overline{\Psi} \right)^2 \delta \phi. 
\end{eqnarray*}
The excitation spectrum is computed by performing the Bogoliubov transformation:
\begin{eqnarray}
\delta \phi &=& \sum_{\alpha} \left[
u_{\alpha} \left( {\bf r} \right) b_{\alpha}
   {\rm e}^{-iE_{\alpha} t/\hbar}
 -
v_{\alpha} \left( {\bf r} \right) b_{\alpha}^{\dagger}
   {\rm e}^{iE_{\alpha} t/\hbar}
\right], \\
\overline{\delta \phi} &=& \sum_{\alpha} \left[
u_{\alpha}^* \left( {\bf r} \right) b_{\alpha}^{\dagger}
   {\rm e}^{iE_{\alpha} t/\hbar}
 -
v_{\alpha}^* \left( {\bf r} \right) b_{\alpha}
   {\rm e}^{-iE_{\alpha} t/\hbar}
\right].
\end{eqnarray}
The energy spectrum $E_{\alpha}$ is calculated by diagonalizing
the skew symmetric matrix, which is carried out by using a routine
in LAPACK.
For the parameter values taken above, we find that
all eigenvalues do not have the imaginary parts within numerical errors.
Therefore, within Gaussian approximation, there is no dynamical instability.
In addition, we find that there is no level crossing.

\section{Summary}
To summarize, we have proposed an experiment to create a quasi-static
sonic horizon using an expanding BEC.
It has been shown that the dynamically formed quasi-static sonic horizon is in hydrodynamic regime
as it should be to discuss analogy with curved spacetime in BEC.
Under suitable choices of the interaction parameter
and the confining potential, 
the characteristic temperature of the particle emission 
is expected to be a few ${\rm nK}$ for sufficiently strong confining potential.
Large number of atoms or strong atomic interaction 
improves the quasi-static nature of the horizon.

Of course, other effect such as cosmological particle creation can occur in this expanding BEC setup, 
as discussed 
in \cite{Barcero02}-\cite{Weinfurtner04}. 
In this paper, we have focused on how to make dynamically formed quasi-static sonic horizon in the hydrodynamic 
regime of the condensate flow. In order to investigate cosmological particle creation
effect and other excitations arising from depletion, we need a different numerical
simulation scheme. The result will be reported in a future publication.

Furthermore, it is interesting to investigate numerically the behavior of
negative frequency modes with positive norm which seem to be related to Hawking effect
as was discussed in \cite{Leonhardt02-1}\cite{Leonhardt02-2}. 
This point shall be investigated in a future publication.

\begin{acknowledgments} 
Y.K. thanks Hideki Ishihara, Ken-ichi Nakao, and  Makoto Tsubota for useful discussions. 
The authors thank Michikazu Kobayashi and Takashi Uneyama for useful 
comments on numerical simulations.
Y.K. was partially supported by the Yukawa memorial foundation.
This work was also supported by 
the 21st Century COE "Center for Diversity and
Universality in Physics" and "Constitution of wide-angle mathematical basis focused on knots" 
from the Ministry of Education, Culture,
Sports, Science and Technology (MEXT) of Japan.
The numerical calculations were carried out on Altix3700 BX2 at YITP 
in Kyoto University.
\end{acknowledgments}

\appendix
\section{Particle creation phenomenon}
\label{particle-creation}

Here we focus on spherically symmetric quantum fluctuations by symmetry.
At $t\leq 0$, the fluid velocity $v_0=0$, and 
the metric of the initial static effective spacetime is 
\begin{eqnarray}
ds^2 \propto  -c_s^2 dt^2 +dr^2+r^2d\Omega_{S^2}^2,
\end{eqnarray} 
where $d\Omega_{S^2}^2$ is the element of solid angle on the unit sphere $S^2$.
After the increase of the interaction, the effective spacetime evolves 
dynamically as the BEC starts to expand.
Then, the sonic horizon is formed as was shown by the above numerical simulation. 
If the effective spacetime is static, 
we can introduce a following time coordinate: 
$\tau = t+ \int v_0 dr/(c_s^2-v_0^2)$, 
and the effective spacetime metric becomes 
\begin{eqnarray}
ds^2 \propto  - (c_s^2-v_0^2 ) d\tau^2 + \frac{c_s^2 dr^2}{c_s^2-v_0^2}+r^2d\Omega_{S^2}^2.
\label{eq:metric2}
\end{eqnarray}
From this expression, it is found that the horizon is located 
at the surface where the condition $c_s=|v_0|$ is satisfied.
A new coordinate $v$ is introduced as $v \equiv \tau+r_*$ where $r_* \equiv \int c_s dr/(c_s^2-v_0^2)$, 
which is a coordinate characterizing ingoing light-like (null) rays in the effective spacetime.

We assume here that the initial state of the quantum field $\varphi$ is 
the vacuum state for the static observer in the initial effective spacetime. 
Under the time evolution of the effective spacetime caused by the expansion of the condensate, 
the creation and annihilation operators for the field $\varphi$ 
also evolve, and particle creation occurs. 

Now we consider an observer who moves along his or her outgoing geodesic 
with proper time $\lambda$, crossing the horizon at $\lambda=0$. 
Hereafter, we term the observer geodesic observer. 
If we assume that the horizon is located at $r=r_H$, 
the proper time $\lambda$ is related to the coordinate $v$ there 
via $ \lambda\approx -\lambda_0 e^{-\frac{2c_H}{\alpha}v}$,
where $c_H \equiv c_s(r_H)$, $\alpha \equiv 2c_H \partial_r(v_0-c_s)|_{r=r_H}$ and $\lambda_0$ is a constant. 
The ingoing mode functions $\varphi_{\omega} = e^{-i\omega v}$ 
have $\lambda$-dependence near the horizon as
\begin{eqnarray}
\varphi_{\omega} \approx \exp\left(i\frac{2c_H \omega}{\alpha}\ln(-\lambda)\right). 
\end{eqnarray}
Initially, the state is the vacuum for the static observer 
and therefore the geodesic observer would see no excitation at short distance, 
because there will be no much higher positive frequency excitations 
than those determined by the time scale of the dynamical expansion of the BEC. 
If we ignore the short distance cut-off determined by the healing length, or equivalently, 
if the latter condition in Eq.(\ref{eq:condition-healing}) is ignored, 
this $\lambda$-dependence of the ingoing mode functions implies that 
the particle creation from the horizon into the inside of the condensate has thermal spectrum 
with the temperature given by (\ref{eq:Hawking-formula}).
Furthermore, even if the short distance cut-off is taken into account,
it is known that the result does not change in principle \cite{Corley:1996ar}.

Therefore, the particle emission from horizon will occur in the case of expanding 
condensate, even where the subsonic region is inside of the horizon.


\end{document}